\documentclass[prl,twocolumn,showpacs,superscriptaddress, floatfix, reprint,longbibliography,nofootinbib]{revtex4-1}

\usepackage[OT1]{fontenc}
\usepackage[usenames,dvipsnames]{color}
\usepackage[latin1]{inputenc}
\usepackage[english]{babel}
\usepackage{graphicx}
\usepackage{color}
\usepackage{amssymb,amsmath}
\usepackage[Gray,squaren]{SIunits}
\usepackage{xspace}
\usepackage{upgreek}
\usepackage{ulem}
\usepackage{epstopdf}
\usepackage{hyperref}
\usepackage{bm}
\newcommand{\fig}[1]{Fig.~\ref{fig:#1}}

\makeatletter
\newcommand*{\balancecolsandclearpage}{%
  \close@column@grid
  \clearpage
  \twocolumngrid
}

\begin{document}

\title{Non-Abelian adiabatic geometric transformations in a cold Strontium gas}
\author{F. Leroux}
\affiliation{Centre for Quantum Technologies, National University of Singapore, 117543 Singapore, Singapore.}
\affiliation{MajuLab, CNRS-UCA-SU-NUS-NTU International Joint Research Unit, Singapore.}
\author{K. Pandey}
\affiliation{Centre for Quantum Technologies, National University of Singapore, 117543 Singapore, Singapore.}
\author{R. Rehbi}
\affiliation{Centre for Quantum Technologies, National University of Singapore, 117543 Singapore, Singapore.}
\affiliation{MajuLab, CNRS-UCA-SU-NUS-NTU International Joint Research Unit, Singapore.}
\author{F. Chevy}
\affiliation{Laboratoire Kastler Brossel, ENS-PSL University, CNRS, Sorbonne Université, Collège de France, 24 rue Lhomond Paris France.}
\author{C. Miniatura}
\affiliation{MajuLab, CNRS-UCA-SU-NUS-NTU International Joint Research Unit, Singapore.}
\affiliation{Centre for Quantum Technologies, National University of Singapore, 117543 Singapore, Singapore.}
\affiliation{Department of Physics, National University of Singapore, 2 Science
Drive 3, Singapore 117542, Singapore.}
\affiliation{School of Physical and Mathematical Sciences, Nanyang Technological University, 637371 Singapore, Singapore.}
\author{B. Gr\'{e}maud}
\affiliation{MajuLab, CNRS-UCA-SU-NUS-NTU International Joint Research Unit, Singapore.}
\affiliation{Centre for Quantum Technologies, National University of Singapore, 117543 Singapore, Singapore.}
\affiliation{Department of Physics, National University of Singapore, 2 Science
Drive 3, Singapore 117542, Singapore.}
\author{D. Wilkowski}
\email{david.wilkowski@ntu.edu.sg}
\affiliation{MajuLab, CNRS-UCA-SU-NUS-NTU International Joint Research Unit, Singapore.}
\affiliation{Centre for Quantum Technologies, National University of Singapore, 117543 Singapore, Singapore.}
\affiliation{School of Physical and Mathematical Sciences, Nanyang Technological University, 637371 Singapore, Singapore.}

\begin{abstract}
{\bf Abstract.} Topology, geometry, and gauge fields play key roles in quantum physics as exemplified by fundamental phenomena such as the Aharonov-Bohm effect, the integer quantum Hall effect, the spin Hall, and topological insulators. The concept of topological protection has also become a salient ingredient in many schemes for quantum information processing and fault-tolerant quantum computation. The physical properties of such systems crucially depend on the symmetry group of the underlying holonomy. Here, we study a laser-cooled gas of strontium atoms coupled to laser fields through a 4-level resonant tripod scheme. By cycling the relative phases of the tripod beams, we realize non-Abelian SU(2) geometrical transformations acting on the dark states of the system and demonstrate their non-Abelian character. We also reveal how the gauge field imprinted on the atoms impact their internal state dynamics. It leads to a thermometry method based on the interferometric displacement of atoms in the tripod beams.
\end{abstract}



\maketitle

\begin{center}\textbf{Introduction}\end{center}

In 1984, M. V. Berry published the remarkable discovery that cyclic parallel transport of quantum states causes the appearance of geometrical phase factors \cite{berry1984quantal}. His discovery, along with precursor works \cite{pancharatnam1956generalized,PhysRev.115.485}, unified seemingly different phenomena within the framework of gauge theories \cite{simon1983holonomy,GeoPhases89}. This seminal work was rapidly generalized to non-adiabatic and noncyclic evolutions \cite{GeoPhases89} and, most saliently for our concern here, to degenerate states by F. Wilczek and A. Zee \cite{wilczek1984appearance}. In this case, the underlying symmetry of the degenerate subspace leads to a non-Abelian gauge field structure. These early works on topology in quantum physics have opened up tremendous interest in condensed matter \cite{baibich1988giant,kato2004observation,konig2007quantum,hsieh2008topological,chang2013experimental} and more recently in ultracold gases \cite{hadzibabic2006berezinskii,lin2011spin,aidelsburger2014measuring,jotzu2014experimental,mancini2015observation,phuc2015controlling,wu2016realization,song2016spin,li2017stripe} and photonic devices \cite{wang2009observation,kuhl2010dirac,schine2016synthetic}.

Moreover, it has been noted that geometrical qubits are resilient to certain noises, making them potential candidates for fault-tolerant quantum computing \cite{zanardi1999holonomic,jones2000geometric,duan2001geometric,solinas2012stability}. So far, beside some recent proposals \cite{kowarsky2014non,sjoqvist2016nonadiabatic}, experimental implementations have been performed for a 2-qubit gate on NV-centers in diamond \cite{zu2014experimental} and for a non-Abelian single qubit gate in superconducting circuits \cite{abdumalikov2013experimental}. These experiments were performed following a non-adiabatic protocol allowing for high-speed manipulation \cite{zhu2002implementation,1367-2630-14-10-103035,sjoqvist2016nonadiabatic}. Recently, coherent control of ultracold spin-1 atoms confined in optical dipole traps was used to study the geometric phases associated with singular loops in a quantum system \cite{bharath2018}. If non-adiabatic manipulations are promising methods for quantum computing, they prevent the study of external dynamic of quantum system in a non-Abelian gauge field, where non-trivial coupling occurs between the internal qubit state dynamics and the center-of-mass motion of the particle.

Here, we report on non-Abelian adiabatic geometric transformations implemented on a non interacting cold fermionic gas of Strontium-87 atoms by using a 4-level resonant tripod scheme set on the $^1S_0, F_g=9/2\rightarrow\,^3P_1, F_e=9/2$ intercombination line at $\lambda = 689\,$nm (linewidth: $\Gamma=2\pi\times 7.5\,$kHz).
About $10^5$ atoms are loaded in a crossed optical dipole trap, optically pumped in the stretched Zeeman state $|F_g=9/2,m_g=9/2\rangle$ and Doppler-cooled down to temperatures $T \sim 0.5\,\mu K$ \cite{chalony2011doppler,Yang2015}, see Methods. A magnetic bias field isolates a particular tripod scheme in the excited and ground Zeeman substate manifolds. Our laser configuration consists of two co-propagating beams (with opposite circular polarizations) and a third linearly-polarized beam orthogonal to the previous ones. These three coplanar coupling laser beams are set on resonance with their common excited state $|e\rangle = | F_e=9/2, m_e=7/2\rangle$.

\begin{figure}
   \begin{center}
      \includegraphics[width =\linewidth]{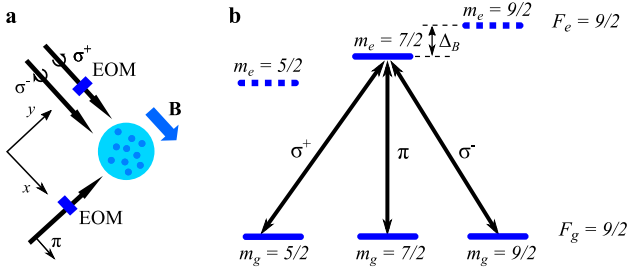}
      \caption{\textbf{Tripod scheme}. (\textbf{a}) Propagation directions of the laser beams and their polarizations along a magnetic bias field $\textbf{B}$. Two Electro-Optics Modulators (EOM) are used to sweep the two independent relative phases of the laser beams. (\textbf{b}) Bare energy level structure of the tripod scheme, implemented on the intercombination line of Strontium-87. The magnetic bias field shifts consecutive excited levels by $\Delta_B\simeq 760\Gamma=5.7\,$MHz. It allows each tripod polarized laser beam to selectively address one of the magnetic transitions marked by the black arrows.}
      \label{fig:1}
   \end{center}
\end{figure}

\begin{center}\textbf{Results}\end{center}
{\bf Dark states basis.} For any value of the amplitude and phase of the laser beams, the effective Hilbert space defined by the four coupled bare levels contains two bright states and two degenerate dark states $|D_1\rangle$ and
$|D_2\rangle$. These dark states do not couple to the excited state $|e\rangle$ and are thus protected from spontaneous emission decay by quantum interference. For equal Rabi transition frequencies, we conveniently choose
\begin{align}
\label{eq:dark2bare}
 |D_1\rangle & =\frac{e^{-i\Phi_{13}({\mathbf r})}|1\rangle-e^{-i\Phi_{23}({\mathbf r})}|2\rangle}{\sqrt{2}}\nonumber\\
 |D_2\rangle & =\frac{e^{-i\Phi_{13}({\mathbf r})}|1\rangle+e^{-i\Phi_{23}({\mathbf r})}|2\rangle-2|3\rangle}{\sqrt{6}},
\end{align}
where $|i\rangle\equiv |m_g\!\!=\!\!i+\!3/2\rangle$ ($i=1,2,3$). $\Phi_{ij}=\Phi_{i}-\Phi_{j}$, where the space-dependent laser phases read $\Phi_i({\mathbf r}) = {\mathbf k}_i \cdot {\mathbf r} + \vartheta_i$. ${\mathbf k}_i$ is the wavevector of the beam coupling state $|i\rangle$ to $|e\rangle$ and $ \vartheta_i$ its phase at origin, see Fig. \ref{fig:1}. To implement non-Abelian transformations on the system, the two independent  offset phases tuned by the Electro-Optic Modulators (EOM), shown in \fig{1}a, are $\phi_i = \vartheta_i-\vartheta_3$ ($i=1,2$).

In a first set of experiments, we probe and quantify the thermal decoherence of the dark states induced by the finite temperature of our atomic sample. In a second set of experiments, we analyze the non-Abelian character of geometric transformations within the dark-state manifold. To do so, we consider a certain phase loop in the parameter space defined by the two relative phases $\phi_1$ and $\phi_2$ of the tripod lasers, and we compare the final populations of the internal atomic states when the cyclic sequence is performed, starting from two different initial points on the loop. In all experiments, we monitor the subsequent manipulation and evolution of the atomic system in the dark-state manifold by measuring the bare ground-state populations with a nuclear spin-sensitive shadow imaging technique, see Methods.

\begin{figure}
   \begin{center}
      \includegraphics[width =\linewidth]{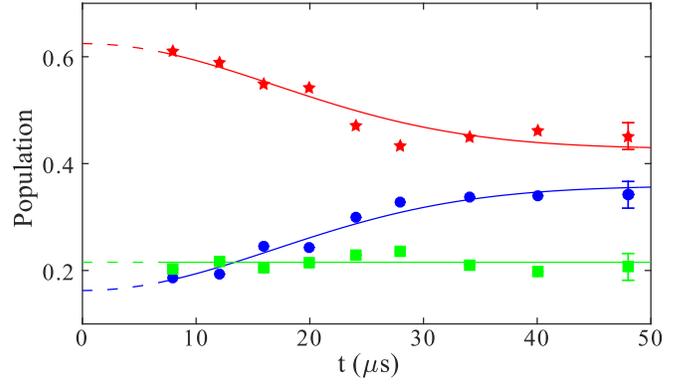}
      \caption{\textbf{Ballistic expansion}. Time evolution of the bare state populations after the tripod ignition sequence (duration $t_0 \simeq 8\,\mu$s) is completed and laser beams have reached equal Rabi frequencies $\Omega=2\pi\times250\,$kHz. The blue circles, the green squares, and the red stars correspond to the populations $\bar{P}_1$, $\bar{P}_2$, and $\bar{P}_3$ with $|i\rangle\equiv | m_g=i+3/2\rangle$ ($i=1,2,3)$, respectively. The error bars correspond to a 95$\%$ confidence interval. Solid lines: theoretical predictions given by equations~\eqref{eq:pop}. The temperature $T$, the initial and the final populations of each spin state are the fit parameters. The dashed lines, at early times, extrapolate the fits into the time window $t_0$. We get a temperature $T= 0.5(1)\,\mu$K meaning that the atoms do not move significantly during the dark state preparation sequence since $\bar{v}t_0/\lambda \simeq 0.08$.}
      \label{fig:2}
   \end{center}
\end{figure}

{\bf Thermal decoherence.} Starting from state $|3\rangle$, we prepare the atoms in dark state $|D_2\rangle$ after a suitable adiabatic laser ignition sequence, see Methods.
We assume that the atoms do not move significantly during the time duration of this sequence, see \fig{2}. Following \cite{ruseckas2005non,PhysRevLett.100.200405}, the subsequent evolution of the atoms is described by the Hamiltonian
\begin{equation}
H=\frac{({\bf \hat{p}} \openone-{\mathbf A})^2}{2M}+W
\label{eq:Scho}
\end{equation}
where ${\bf \hat{p}} = -i \hbar \boldsymbol \nabla$ is the momentum operator, $\openone$ is the identity operator in the internal dark-state manifold, $M$ the atom mass, $\mathbf A$ the geometrical vector potential with matrix entries  $\mathbf A_{jk} = i\hbar\langle D_j|\boldsymbol\nabla D_k\rangle$ and $W$ the geometrical scalar potential with matrix entries
\begin{equation}
W_{jk}=\frac{\hbar^2\langle\boldsymbol\nabla D_j|\boldsymbol\nabla D_k\rangle- \mathbf (\mathbf A^2)_{jk}}{2M}.
\label{eq:W}
\end{equation}

With our laser geometry, $\mathbf A$, $\mathbf A^2$, and $W$ have the same matrix form, and are uniform and time-independent, see Methods. Thus, we can look for states in the form
$|\psi\rangle \otimes |\mathbf p\rangle $ where $\mathbf p=M \mathbf v$ is the initial momentum of the atoms and $|\psi\rangle$ some combination of dark states. Denoting by $P_0(\mathbf v)$ the initial atomic velocity distribution, we find that the population of state $|2\rangle$ remains constant while the two others display an out-of-phase oscillatory behavior at a velocity-dependent frequency $\omega_v = \frac{2}{3}[k(v_x-v_y)+2\omega_R]$:
\begin{equation}
 \begin{aligned}
  P_1(\mathbf v,t)&=\frac{5P_0(\mathbf v)}{12} \left(1 - \frac{3}{5} \cos\omega_v t\right),\\
  P_2(\mathbf v,t)&=\frac{P_0(\mathbf v)}{6},\\
  P_3(\mathbf v,t)&=\frac{5P_0(\mathbf v)}{12} \left(1 + \frac{3}{5} \cos\omega_v t\right),
 \end{aligned}
\label{eq:Pv}
\end{equation}
where $\omega_R = \hbar k^2/(2M)$ is the recoil frequency and $k=2\pi/\lambda$ is the laser wavenumber. The frequency component proportional to $k(v_x-v_y)$ comes from the momentum-dependent coupling term ${\mathbf A} \cdot {\bf {\hat p}}/M$ in equation~\eqref{eq:Scho} (Doppler effect) whereas the other frequency component, proportional to $\omega_R$, comes from the scalar term ${\mathbf A}^2/(2M)+W$. With our laser configuration, light-assisted mechanical forces can only come from photon absorption and emission cycles between a pair of orthogonal laser beams. Such photon exchanges would induce a population change of state $|2\rangle$. Since no force is acting here on the centre-of-mass of the atoms (the Abelian gauge field is uniform and can be gauged away), the population $P_2(\mathbf v, t)$ must stay constant, as predicted by equation~\eqref{eq:Pv}. Since photon absorption and emission cycles between the pair of co-propagating laser beams do not impart any net momentum transfer to the atoms, population transfer between states $|1\rangle$ and $|3\rangle$ is possible and $P_1(\mathbf v, t)$ and $P_3(\mathbf v, t)$ change in time, their sum being constant due to probability conservation.

Averaging over the Maxwellian velocity distribution of the atoms, the bare state populations of the thermal gas read
\begin{align}
\bar{P}_{1}(t) & = \frac{5}{12} - \frac{1}{4}\cos{\left(\frac{4}{3}\omega_Rt\right)}\exp{\left[-\frac{4}{9}\left(k\bar{v}t\right)^{2}\right]},\nonumber\\
\bar{P}_2(t) & = \frac{1}{6},\nonumber\\
\bar{P}_3(t) & = \frac{5}{12} + \frac{1}{4}\cos{\left(\frac{4}{3}\omega_Rt\right)}\exp{\left[-\frac{4}{9}\left(k\bar{v}t\right)^{2}\right]},
\label{eq:pop}
\end{align}
where $\bar{v}=\sqrt{k_BT/M}$ is the thermal velocity of the gas at temperature $T$. We see that $\bar{P}_1$ and $\bar{P}_3$ converge to the same value at long times. This means that the thermal average breaks the tripod scheme into a $\Lambda$-scheme coupled to the two circularly-polarized beams and a single leg coupled to the linearly-polarized beam. As a consequence, quantum coherence partially survives the thermal average.

Our experimental results confirm this behavior even if $\bar{P}_1$ and $\bar{P}_3$ do not merge perfectly, see Fig. \ref{fig:2}. This discrepancy can be lifted by introducing a $10\%$ imbalance between the Rabi transition frequencies in our calculation. The population difference $\bar{P}_3-\bar{P}_1$ measures in fact the Fourier transform of the velocity distribution along the diagonal direction ${\bf \hat{x}}-{\bf \hat{y}}$. It decays with a Gaussian envelope characterized by the time constant $\tau=3/(2k\bar{v})$, as predicted by equations~\eqref{eq:pop}. This interferometric thermometry is similar to some spectroscopic ones such as recoil-induced resonance \cite{courtois1994recoil,PhysRevA.50.R1992} or stimulated two-photons transition \cite{PhysRevLett.66.2297,PhysRevA.85.063416}. From our measurements, we get $T= 0.5(1)\,\mu$K, $\tau \simeq 24\,\mu$s and $\bar{v}\simeq 6.9\,$mm/s.

\begin{figure}
   \begin{center}
      \includegraphics[width =0.95\linewidth]{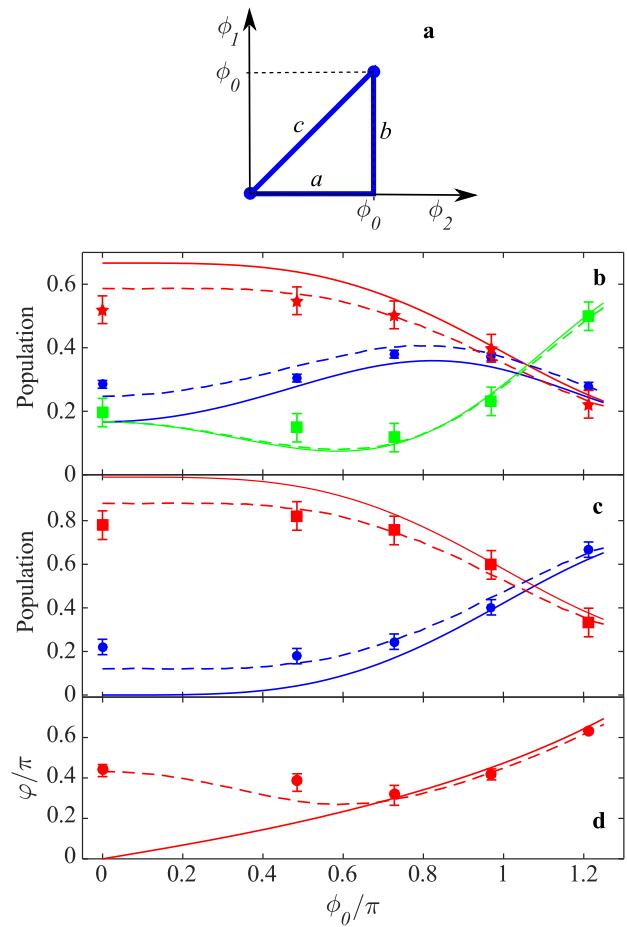}
      \caption{\textbf{Geometric gate operation}. (\textbf{a}) Phase loop in parameter space ($\phi_2$, $\phi_1$). $\phi_i$ ($i=1,2$) are the two independent offset phases tuned by the EOMs shown in \fig{1}a. We have performed two counterclockwise cycles: the first one is $a \to b \to c$ and starts from the origin, the second one is $c \to a \to b$ and starts from the upper corner. The loop is completed in $12\,\mu$s and its excursion is $\phi_0$. (\textbf{b}) Measured bare state populations $\bar{P}_1$ (blue circles), $\bar{P}_2$ (green squares) and $\bar{P}_3$ (red stars) as a function of $\phi_0$ for the first cycle. The Rabi frequencies are $\Omega=2\pi\times 450\,$kHz and $T=0.5\,\mu$K. Dark state reconstruction as a function of $\phi_0$. (\textbf{c}) Population of $|D_1\rangle$ (blue circles) and $|D_2\rangle$ (red squares). (\textbf{d}) Azimuthal phase $\varphi$.
      The solid and dashed curves in panels (\textbf{b}), (\textbf{c}), and (\textbf{d}) are the theoretical predictions for a pinned atom and for a gas at temperature $T$ respectively. The error bars correspond to a 95$\%$ confidence interval.
      }
      \label{fig:3}
   \end{center}
\end{figure}

{\bf Non-Abelian transformations.} We now investigate the geometric non-Abelian unitary operator $U$ acting on the dark-state manifold when the relative phases of the tripod beams are adiabatically swept along some closed loop $\mathcal{C}$ in parameter space. For a pinned atom ($M\rightarrow\infty$), $U$ is given by the loop integral along $\mathcal{C}$ of the $2\times 2$ Mead-Berry 1-form $\omega\equiv[\omega_{jk}]\equiv[i\hbar\langle D_j|d D_k\rangle]$
\begin{equation}
   U=\mathfrak{P}\exp{\left(\frac{i}{\hbar}\oint_\mathcal{C} \omega\right)},
\label{eq:U}
\end{equation}
where $\mathfrak{P}$ is the path-ordering operator \cite{wilczek1984appearance}.

As before, the system is initially prepared in dark state $|D_2\rangle$. Then, starting from the origin, the phase loop is cycled counterclockwise, see Fig. \ref{fig:3}a.
Each segment is linearly swept in $\Delta t=4\,\mu$s and the phase excursion is $\phi_0$. The total duration $3\Delta t$ of the loop is thus less than the thermal decoherence time $\tau$ discussed above. In Fig. \ref{fig:3}b, we plot the bare state populations measured right after the phase loop as a function of $\phi_0$ and their comparison to theoretical predictions for pinned atoms and for atoms at finite temperature under the adiabatic assumption. This clearly shows that thermal effects are an important ingredient to reproduce the experimental results and that the adiabatic approximation is well justified, see Methods.
Note that the mismatch with pinned atoms decreases with increasing $\phi_0$.
This is because the thermal decoherence is quenched by the increasing geometrical coupling among the dark states when the sweep rate $\gamma =\phi_0/\Delta t > k\bar{v}, \omega_R$ \footnote{The thermal decoherence quenching can be quantified by the bare population distance $\Delta P=\sqrt{\sum_{i=1}^3(P_i-P_{0i})^2}$, where $P_i$ and $P_{0i}$ are the experimental and pinned-atom populations. At $\phi_0=\pi$, we get $\Delta P=0.04(5)$. This value increases when $\phi_0$ decreases, reaching  $\Delta P=0.19(5)$ at $\phi_0=0$}. As a further approximation, we now disregard thermal decoherence and consider that the system after the phase loop is described by a pure quantum state $|\psi_{out}\rangle=\sum_{j=1,2}d_j|D_j\rangle$ \footnote{A pure state is denoted by a density matrix $\rho$ fulfilling Tr$\{\rho^2\}=1$. For a finite-temperature gas, we find Tr$\{\rho^2\}= 0.95$ at $\phi_0=\pi$. This value decreases when $\phi_0$ decreases, reaching Tr$\{\rho^2\}=0.8$ at $\phi_0=0$}.
As shown in Figs. \ref{fig:3}b and \ref{fig:3}c, one can easily extract the dark state populations and the absolute value of the azimuthal angle $\varphi=\textrm{Arg}(d_2)-\textrm{Arg}(d_1)$ from the measured bare state populations, see Methods. When $\phi_0 \gtrsim 0.7\pi$, the values for $\varphi$ match well with the prediction for a pinned atom confirming the quenching of thermal decoherence. At $\phi_0=\pi$, the two dark state populations are almost equal. In the language of the Bloch sphere representation, this corresponds to a rotation of the initial south pole state $|D_2\rangle$ to the equatorial plane.

\begin{figure}
   \begin{center}
      \includegraphics[width =0.45\linewidth, angle=90]{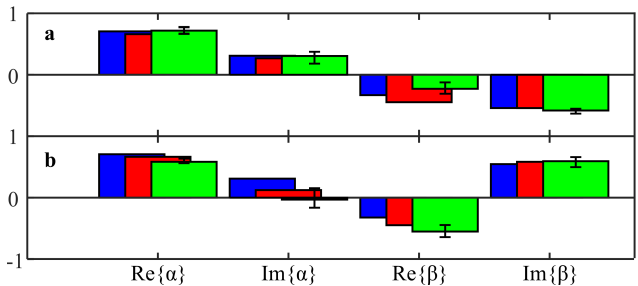}
      \caption{\textbf{Unitary operators reconstruction}. (\textbf{a}) Elements of operator $U$ given by equation~\eqref{paramU} for
      the phase loop $a \to b \to c$ at $\phi_0=\pi$, see Fig.~\ref{fig:3}a. The blue, red, and green bars correspond to a pinned atom, a gas at temperature $T=0.5\,\mu$K, and the experimental data respectively. (\textbf{b}) Same as (\textbf{a}) but for the phase loop $c \to a \to b$. The error bars correspond to a 95$\%$ confidence interval.
      }
      \label{fig:4}
   \end{center}
\end{figure}

We now reconstruct the full geometric unitary operator $U$ for $\phi_0=\pi$. Up to an unobservable global phase, we write:
\begin{equation}
\label{paramU}
U=
\begin{bmatrix}
   \alpha       &  \beta \\
   -\beta^*       &  \alpha^*
\end{bmatrix}
\end{equation}
with $|\alpha|^2+|\beta|^2=1$. The previous dark state reconstruction, done after the phase loop applied on $|D_2\rangle$, gives access to $|\alpha|$, $|\beta|$ and $\textrm{Arg}(\alpha)-\textrm{Arg}(\beta)$, see Methods. To obtain $\textrm{Arg}(\alpha)$ and $\textrm{Arg}(\beta)$ and fully determine $U$, we start from a linear combination of dark states $|D_1\rangle$ and $|D_2\rangle$, perform the phase loop and process the new data. The results are shown in Fig. \ref{fig:4}a and compared to the theoretical predictions for a pinned atom and a gas at finite temperature. The good agreement with our data validates the expected small impact of temperature for $\phi_0=\pi$.

{\bf Probing non-Abelianity.} With the previous phase loop protocol, we have $U=U_cU_bU_a$ where $a$, $b$, and $c$ label the edges of the loop, see Fig.~\ref{fig:3}a. To illustrate the non-commutative nature of the transformation group, we will cycle the phase loop counterclockwise starting from the upper corner. We then reconstruct the corresponding unitary operator $U'=U_bU_aU_c$ like done for $U$. The results are depicted in Fig. \ref{fig:4}b and show that $U$ and $U'$ are indeed different, though unitarily related, confirming the sensitivity of these geometric transformations to path ordering. The Frobenius distance between the two unitaries is $D=\sqrt{2-|{\rm Tr}(U^\dag U')|}=1.27(25)$ and is in agreement with the theoretical result for a pinned atom ($D=1.09$) and for a finite-temperature gas ($D=1.14$). These values have to be contrasted with the maximum possible Frobenius distance $D=2$.

\begin{center}\textbf{Discussion}\end{center}

Using a tripod scheme on Strontium-87 atoms, we have implemented adiabatic geometric transformations acting on two degenerate dark states. This system realizes a universal geometric single-qubit gate. We have studied SU(2) transformations associated to laser beams phase loop sequences and
shown their non-Abelian character. In contrast to recent works done in optical lattices \cite{aidelsburger2014measuring,jotzu2014experimental,mancini2015observation,phuc2015controlling,wu2016realization,song2016spin,li2017stripe}, our system realizes an artificial gauge field in continuous space.
Depending on the laser field configuration, different manifestations of artificial gauge fields can be engineered such as spin-orbit coupling \cite{Jacob2007,PhysRevLett.100.200405}, Zitterbewegung \cite{PhysRevLett.100.200405}, magnetic monopole \cite{ruseckas2005non} or non-Abelian Aharomov-Bohm effect \cite{Jacob2007} (see \cite{dalibard2011colloquium,goldman2014light} for reviews). A generalization to the SU(3) symmetry is also discussed in \cite{hu2014u}. Some of these schemes might be difficult to implement in optical lattices. Gauge fields generated by optical fields come from a redistribution of photons among the different plane waves modes and involve momenta transfer
comparable to the photon recoil. Observing mechanical effects of non-uniform or non-Abelian gauge fields would thus require atomic gases colder than the recoil temperature and thus cooling techniques beyond the mere Doppler cooling done here \cite{desalvo2010degenerate,tey2010double}. However, the gauge field is still driving the internal state dynamics  regardless of the temperature of the gas provided the adiabatic condition is fulfilled. Noticeably, this internal state dynamics is still present when the gauge field is Abelian and uniform. It led us to an interferometric thermometry based on the Fourier transform of the velocity distribution of the gas.

\begin{center}\textbf{Methods}\end{center}

\begin{figure}
   \begin{center}
      \includegraphics[width =1\linewidth]{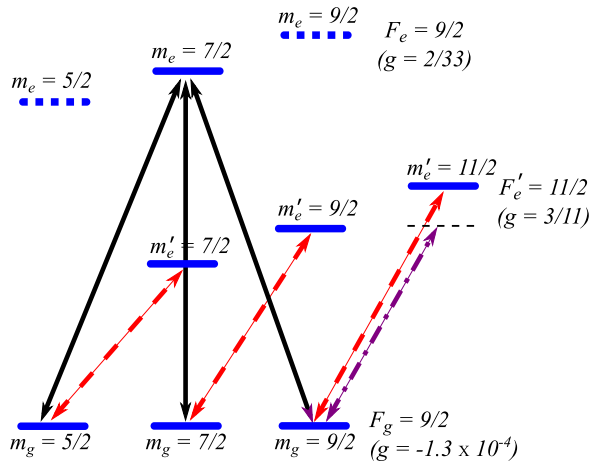}
      \caption{\textbf{Energy levels and experimentally relevant transitions}. A magnetic bias field $B=67$\,G lifts the degeneracy of the different Zeeman manifolds and allows to address each transition individually. The Land\'{e} factors $g$ are indicated for each hyperfine level. The black arrows correspond to the tripod beams (see main text for more details). The dashed red arrows indicate the transitions used for the shadow spin-sensitive imaging system. The dash-dotted purple arrow is the red-detuned cooling transition used in the far-off resonant dipole trap. }
      \label{fig:FidS1}
   \end{center}
\end{figure}

{\bf Cold sample preparation and implementation of the tripod scheme.} The cold gas is obtained by laser cooling on the $^1S_0\rightarrow\,^3P_1$ intercombination line at $689\,$nm (linewidth $\Gamma=2\pi\times 7.5\,$kHz). Atoms are first laser cooled in a magneto-optical trap and then transferred into an ellipsoidal crossed optical dipole trap at $795$\,nm (trapping frequencies $150$, $70$, and $350$ Hz) where they are held against gravity. Atoms are then optically pumped in the stretched $m_g=F_g=9/2$ magnetic substate and subsequently Doppler cooled in the optical trap using the close $m_g=F_g=9/2\rightarrow m'_e=F'_e=11/2$ transition, see Fig. \ref{fig:FidS1}. The atomic cloud contains about $10^5$ atoms at a temperature $T=0.5\,\mu$K (recoil temperature $T_R = \hbar\omega_R/k_B \approx 0.23\,\mu$K, where $k_B$ is the Boltzmann constant). A magnetic field bias of $B=67$\,G is applied to lift the degeneracy of the Zeeman excited states. Because the Zeeman shift between levels in the excited manifold $F_e=9/2$ is large, one can isolate a tripod scheme between three ground-state levels and a single excited state, namely $|e\rangle = |F_e=9/2, m_e=7/2\rangle$, as indicated in Fig. \ref{fig:FidS1}. The Zeeman shift of the ground-state levels (Landé factor $g=-1.3\times 10^{-4}$) is weak (12 kHz) and is compensated by changing accordingly the frequencies of the three tripod laser beams. The lasers are finally tuned at resonance and their polarizations are chosen according to the electrical dipole transition selection rules. In practice, the two laser beams with right and left circular polarizations, respectively addressing the $m_g=5/2\rightarrow m_e=7/2$ and $m_g=9/2\rightarrow m_e=7/2$ transitions, are co-propagating. The laser beam with linear polarization, aligned with
the magnetic bias field, addressing the $m_g=7/2\rightarrow m_e=7/2$,  is orthogonal to the circularly polarized beams, see \fig{1}b. The plane of the lasers is chosen orthogonal to the direction of gravity. The two independent laser offset phases $\phi_1$ and $\phi_2$ (see main text) can be tuned by using two electro-optic modulators. \\

{\bf Adiabatic approximation.} The two independent laser offset phases $\phi_1$ and $\phi_2$ are ramped from $0$ to $\phi_0 \leq 1.2\pi$ at a constant rate $\gamma$ during the sweep time $\Delta t=4\,\mu$s. The AC-Stark shifts of the bright states is given by $\sqrt{3}\Omega=2\pi\times 780\,$kHz. Since $\gamma=\phi_0/\Delta t \leq 2\pi\times 150\,$kHz, we have $\sqrt{3}\Omega/\gamma \geq 5.2$ and the adiabatic approximation is well justified.\\

{\bf Initial dark state preparation.} Starting with atoms in the $|m_g=9/2\rangle$ stretched state, the tripod beams are turned on following two different sequences. The first sequence prepares dark state $|D_2\rangle$, see equation~\eqref{eq:dark2bare}. More precisely, we first turn on the two laser beams connecting the empty bare states $|m_g=5/2\rangle$ and $|m_g=7/2\rangle$ to the excited state $|m_e = 7/2\rangle$ and then adiabatically ramp on the last laser beam. This projects state $|m_g=9/2\rangle$ onto $|D_2\rangle$ with a fidelity of $95\%$. Since the bare state $|m_g=9/2\rangle$ is only present in $|D_2\rangle$, our choice of basis in the dark-state manifold is well adapted to understand the dark state preparation. A different ignition sequence is used to prepare a combination of dark states $|D_1\rangle$ and $|D_2\rangle$. By turning on sequentially abruptly the left-circular beam, and adiabatically the right-circular beam, we create a coherent (dark) superposition of the state $|m_g=5/2\rangle$ and $|m_g=9/2\rangle$. Finally, we turn on abruptly the linearly-polarized beam and we expect to produce the linear combination $(|D_1\rangle+\sqrt{3}|D_2\rangle)/2$. In practice, a systematic  phase rotation occurs once the last beam is turned on which adds an extra mixing among the dark states. Performing the bare state population analysis, we find that this initial state corresponds in fact to $0.6 |D_1\rangle+0.8 e^{i 0.15\pi} |D_2\rangle$.\\

{\bf Spin sensitive imaging system.} The bare state populations in the ground-state are obtained with a nuclear spin-sensitive shadow imaging technique on the $F_g=9/2\rightarrow F'_e=11/2$ line, see Fig. \ref{fig:FidS1}. First we measure the population of state $|m_g=9/2\rangle$ with a shadow laser tuned on the closed $m_g=9/2\rightarrow m'_e=11/2$ transition. Then, using the same atomic ensemble, we measure the population of state $|m_g=7/2\rangle$ by tuning the shadow laser on the $m_g=7/2\rightarrow m'_e=9/2$ transition. This transition is open but its large enough Clebsch-Gordan coefficient ($\sqrt{9/11} \sim 0.9$) ensures a good coupling with the shadow laser. The population of state $|m_g=5/2\rangle$ is measured in the same way ($m_g=5/2\rightarrow m'_e=7/2$ open transition, its Clebsch-Gordan coefficient $\sqrt{36/55}\sim 0.8$ being still large enough). The shadow laser beam shines the atoms during $40\,\mu$s with an on-resonance saturation parameter $I/I_s = 0.5$ (saturation intensity $I_s = 3 \, \mu {\rm W/cm}^2$). With such values, the average number of ballistic photons scattered per atom is less than one and optical pumping can be safely ignored, ensuring an accurate measurement of the ground-state populations. To achieve a good statistics, the same experiment was repeated 100 times and the corresponding data averaged. The error bars on the bare state populations correspond to a 95$\%$ confidence interval.\\

{\bf Dark states and unitary matrix reconstruction.} A state in the dark-state manifold takes the form $|\psi\rangle= \sum_{j=1,2} d_j \, |D_j\rangle$ with $|d_1|^2$ and $|d_2|^2=1-|d_1|^2$ the populations of states $|D_1\rangle$ and $|D_2\rangle$ and $\varphi = {\rm Arg}(d_2)-{\rm Arg}(d_1)$ the azimuthal angle. Using equation~\eqref{eq:dark2bare}, we immediately find
\begin{equation}
\begin{aligned}
|d_2| &= \sqrt{3\bar{P}_3 / 2},\nonumber\\
\cos\varphi &= (\bar{P}_1-\bar{P}_2)/\sqrt{\bar{P}_3(2-3\bar{P}_3)}.
\end{aligned}
\end{equation}
Do note that the normalization of $|\psi\rangle$ restricts the possible values of the $\bar{P}_i$ summing up to 1.
The sign of $\varphi$ is determined using the prediction of equation~\eqref{eq:U} for a pinned atom ($M\to\infty$).

To reconstruct the unitary matrix $U$, as expressed in equation~\eqref{paramU}, we perform the phase loop sequence on two different initial dark states (their representative points on the Bloch sphere should not be opposite) and perform the dark state reconstruction for each of them. The two phase terms in $U$ are reconstructed up to a sign. As for the dark state reconstruction, we rely on the prediction for a pinned atom to lift this sign ambiguity. \\

{\bf Gauge fields and adiabatic Schr\"{o}dinger equation.} The time-dependent interaction operator for the resonant tripod scheme, in the rotating-wave approximation, has the following expression:
\begin{equation}
H(t)=\frac{\hbar\Omega(\textbf{r},t)}{2}\sum_{i=1}^3|e\rangle\langle i|+H.c.
\end{equation}
We assume here that the laser Rabi frequencies coupling the ground-states $|i\rangle =|m_g=i+3/2\rangle$ to the excited state $|e\rangle = |m_e=7/2\rangle$ have all the same amplitude denoted by $\Omega$. The time dependency comes from the cyclic ramping sequence of the two offset laser phases $\phi_j$ ($j=1,2$). Neglecting transitions outside the dark-state manifold (adiabatic approximation), the system is described by a quantum state $|\psi(\textbf{r},t)\rangle=\Sigma_{j=1,2}\Psi_j(\textbf{r},t)|D_j(\textbf{r},t)\rangle$, where $\Psi_j$ is the wave function of the centre-of-mass of the atom in an internal state $|D_j\rangle$. In this basis, the adiabatic Schr\"{o}dinger equation for the column vector $\underline{\Psi}=(\Psi_1,\Psi_2)^T$ reads:
\begin{equation}
i\hbar\dot{\underline{\Psi}}=\left[\frac{(\hat{\textbf{p}}\openone-\textbf{A})^2}{2M}+W-\omega_t\right]\underline{\Psi},
\end{equation}
where the dot denotes time derivative. The first two terms on the right-hand side describe the dynamics of an atom subjected to the synthetic gauge field. The last term $\omega_t\equiv[\omega_{jk}]\equiv[i\hbar\langle D_j|\dot{D_k}\rangle]$ is due to the cyclic ramping sequence of the laser phases. Only this term remains for a pinned atom ($M\rightarrow+\infty$), in which case one recovers equation~\eqref{eq:U}. The general expressions of $\textbf{A}$ and $W$ are given in the main text. With equal and constant Rabi frequencies amplitude, and for the orientation of our laser beams, one finds:
\begin{equation}
\begin{aligned}
\textbf{A}&=\frac{2\hbar(\textbf{k}_2-\textbf{k}_1)}{3} \, \mathcal{M},\nonumber\\
\frac{\textbf{A}^2}{2M}&=\frac{8E_R}{9} \, \mathcal{M},\nonumber\\
W&=-\frac{4E_R}{9} \, \mathcal{M},
\end{aligned}
\end{equation}
where $E_R=\hbar\omega_R=\hbar^2k^2/(2M)$ is the recoil energy and $\mathbf k_j$ is the wavevector of laser beam $j$ (see main text). As one can see, all these operators have the same matrix form. The matrix $\mathcal{M}$ reads:
\begin{equation}
\mathcal{M} = \frac{\openone + \mathbf s \cdot \boldsymbol{\sigma}}{2} = \left(
  \begin{array}{cc}
    3/4 & -\sqrt{3}/4 \\
    -\sqrt{3}/4 & 1/4 \\
  \end{array}
\right)
\end{equation}
and satisfies $\mathcal{M}^2= \mathcal{M}$, its unit Bloch vector being $\mathbf s = (-\sqrt{3}/2, 0, 1/2)$. As a consequence, all these operators can be diagonalized at once by the same transformation and amenable to the simple projector matrix form:
\begin{equation}
\mathcal{M} \to \mathcal{M}_D  = \left(
  \begin{array}{cc}
   1 & 0 \\
   0 & 0 \\
  \end{array}
\right).
\end{equation}
Because of our laser beams geometry, the vector potential $\textbf{A}$ is in fact Abelian since its only non-zero matrix component is along $\textbf{k}_2-\textbf{k}_1$.

In contrast, the operator $\omega_t$ has a different matrix form. Indeed, the two offset phases $\phi_j$ of the lasers (see main text) can be addressed at will. Following \cite{dalibard2011colloquium}, we get:
\begin{equation}
\omega_t=\frac{\hbar}{2}\left(
  \begin{array}{cc}
    \dot{\phi}_1+\dot{\phi}_2 & (\dot{\phi}_1-\dot{\phi}_2)/\sqrt{3} \\
    (\dot{\phi}_1-\dot{\phi}_2)/\sqrt{3} & (\dot{\phi}_1+\dot{\phi}_2)/3 \\
  \end{array}
\right).
\end{equation}
In particular, we note that $\omega_t$ leads to non-Abelian transformations. An immediate consequence is that, for a given phase loop in parameter space, the geometric unitary operator associated with a cycle of phase ramps depends on the starting point of the cycle on the loop. Different starting points lead to different, though unitarily related, non-commuting geometric unitary operators.\\

\textbf{Data availability.} The authors declare that the main data supporting the findings of this study are available within the article. Extra data are available from the corresponding author upon request.

\textbf{Acknowledgements.} F. C. thanks CQT and UMI MajuLab for their hospitality. The Authors thank Mehedi Hasan for his critical reading of the manuscript.
C. M. is a Fellow of the Institute of Advanced Studies at Nanyang Technological University (Singapore). This work was supported by the CQT/MoE funding Grant No. R-710-002-016-271.

\textbf{Author contributions.} F. L., K. P., and R. R. have developed the experimental platform and performed the experiments. F. L. and D. W. have analyzed the experimental data. B. G., F. C., C. M., and D. W. have developed the models. All authors have contributed to the writing of the manuscript.

\textbf{Competing interests:} The authors declare non-financial interests.

\newcommand{\noopsort}[1]{}\providecommand{\noopsort}[1]{}\providecommand{\singleletter}[1]{#1}%

\end{document}